# Optimizing traffic signs and lights visibility for the teleoperation of autonomous vehicles through ROI compression

I. Dror, O. Hadar, *Senior Member, IEEE*

**Abstract** Autonomous vehicles are a promising solution to traffic congestion, air pollution, accidents, and wasted time and resources. However, remote driver intervention may be necessary for extreme situations to ensure safe roadside parking or complete remote takeover. In such cases, high-quality real-time video streaming is crucial for practical remote driving. In a preliminary study, we already presented a region of interest (ROI) HEVC data compression where the image was segmented into two categories of ROI and background, allocating more bandwidth to the ROI, yielding an improvement in the visibility of the classes that essential for driving while transmitting the background with lesser quality. However, migrating bandwidth to the large ROI portion of the image doesn't substantially improve the quality of traffic signs and lights. This work categorized the ROIs into either background, weak ROI, or strong ROI. The simulation-based approach uses a photo-realistic driving scenario database created with the Cognata self-driving car simulation platform. We use semantic segmentation to categorize the compression quality of a Coding Tree Unit (CTU) according to each pixel class. A background CTU can contain only sky, trees, vegetation, or building classes. Essentials for remote driving include significant classes such as roads, road marks, cars, and pedestrians.
And most importantly, traffic signs and traffic lights. We apply thresholds to decide if the number of pixels in a CTU of a particular category is enough to declare it as belonging to the strong or weak ROI. Then, we allocate the bandwidth according to the CTU categories. Our results show that the perceptual quality of traffic signs, especially textual signs and traffic lights, improves significantly by up to 5.5 dB compared to the only background and foreground partition, while the weak ROI classes at least retain their original quality.

*Index Terms*— **Self-driving cars, teleoperation, video encoding, HEVC, ROI compression, Semantic Segmentation**

## I. INTRODUCTION

Developing a fully automated self-driving car beyond SAE (Society of Automotive Engineers) Level 3 is a challenging task [1]. Designing a fully self-maneuverable vehicle that can drive safely in any driving conditions, including fog, heavy rain, road obstacles, complex roadside works, unpolite and incapable drivers, and pedestrians on the road is challenging. This problem is much more complicated than designers thought at first.

A fallback solution is to allow remote driver control of the vehicle when the self-driving car's automation encounters difficulty in handling the driving tasks independently. The idea is that most of the time, driving will be autonomous, and only in rare cases, such as extreme weather, very dense driving situations, and obstacles, will the remote driver take control of the vehicle. To make such a system cost-effective, in most remote driver interventions, the distant driver will drive the vehicle shortly to a safe and nearby location, and then support will arrive at the vehicle.

Reliable and safe teleoperation by a remote driver in a distant control room requires high quality, low delay, and reliable video communication. The available cellular 4 and 5G network bandwidth varies significantly when the car moves and switches from one cell to another. Furthermore, varying distances between car and network antennas and network congestion makes a video broadcast over the cellular Network demanding [2].

The required Glass-to-Glass (G2G) delay for the transmission of video is measured as the time passed from the rays of an event hitting the camera's lens until the teleoperator watches the same event on his monitor [3]–[5]. Safe and reliable teleoperation of a self-driving car by a remote driver requires a maximal end-to-end delay of 175 ms. Remote driving leaves approximately 100 ms for the G2G delay. A study of the influence of the delay in a multiparty racing car game shows that a player can respond well if the delay is smaller than 100 ms. Therefore, reliable low-delay video transmission by a teleoperator field is crucial for safe driving [6].

This work proposes an improvement in video quality by applying multi-category ROI compression. Instead of parsing the image to ROI and background only, we use two categories: weak and strong ROI. The most critical objects for driving are the traffic signs and lights, which are assigned much of the relative (to their size) bandwidth, enhancing their visibility while at least maintaining the quality of the weak ROI.

The video encoder compresses video coding blocks at a higher quality if they contain more critical elements for safe driving. Conversely, coding blocks that are less important for safe driving are considered background and compressed with lower quality.

The requirement for fast teleoperator response and low delay video cause the granularity of the RC to be at the level of a single frame. On the other hand, many video applications that are less prone to real-time performance have RC granularity in Group of Pictures (GOP) or time frame (for example, a second) granularities.

The Israel Innovation Authority supported this work through the Andromeda Consortium.

Itai Dror and Ofer Hadar are with the School of Electrical and Computer Engineering, Ben Gurion University of the Negev, Beersheba, Israel. email: (itaidro@post.bgu.ac.il, hadar@bgu.ac.il).

The car's edge computer analyzes the video and decides which parts are more important for remote driving and which are less critical. There are several methods for image parsing and understanding. First, Deep Learning (DL) semantic segmentation tools [7] can classify every image pixel by its class, and then it is possible to assign an image block an ROI category based on the block content. Second, DL object detection methods [7] can be trained to find the bounding boxes of essential objects such as vehicles, traffic lights, and traffic signs. Then, coding blocks are mapped and assigned ROI priorities only if they contain parts of the crucial objects. Third, the car's Lidar (Light detecting and ranging) depth maps [8] can distinguish between ROI and non-ROI coding blocks according to their distances from the vehicle. After mapping the image into non-ROI and ROI coding blocks, the encoder allocates bandwidths to both regions. Finally, it assigns the bitrates according to the temporal bandwidth available from the network operator, the relative sizes of the ROI and non-ROI, and the predefined weight that defines the normalized bitrate ratio of non-ROI to ROI.

This work proposes further improving video quality by applying multi-category ROI compression. Instead of parsing the image to ROI and background only, we use two categories: weak and strong ROI. The most critical objects for driving, which are the traffic signs and lights, are assigned much of the relative (to their size) bandwidth, enhancing their visibility while at least maintaining the quality of all other objects that are also important for the distant driver.

The work relies on several concurrent technologies: self-driving cars, teleoperation of autonomous vehicles when necessary, Complementary metal-oxide-semiconductor (CMOS) video cameras, Lidar sensors, Radars, Global Positioning System (GPS), High-Efficiency Video Coding (HEVC), also known as H.265 video compression standard, Cognata [9] autonomous vehicles photo-realistic simulator, and deep neural networks that can analyze the video image for the following tasks of object classification, object recognition, object tracking, semantic segmentation, instance segmentation and more.

## A. Self-driving cars

Self-driving cars will change our public or private attitude toward transportation, driving, car ownership, car safety, car accidents, and loss of working hours.

The SAE defines five levels of self-driving car automation [10][11]. Level 0 represents a conventional car where the human driver controls the vehicle. Most of the cars that are available today are of Level 0. However, the human driver functionalities are migrating into the car's automation as the SAE level increases. At Level 1, the vehicle has driving assistance such as automatic braking, lane assistance, and adaptive cruise control. The car's automation can control the car's speed or the steering, but not both. The driver is responsible for monitoring the road and must take over immediately if the assistance system fails.

Level 2 is for partial automation. The car can control the steering, accelerate, and brake if possible. The human driver is responsible for more complicated tasks such as changing lanes, responding to traffic signals, and looking for hazardous conditions. The driver's hands should take control of his vehicle immediately when requested. Examples of Level 2 systems are Tesla autopilot and Mercedes-Benz Driver Assistance Systems. From Level 3 and up, car automation can take over in certain traffic conditions, such as assistance in traffic jams or acting as a highway chauffeur. At Level 3, the driver can do other tasks than driving. Texting on a cell phone, reading, writing, and even relaxing are possible. The driver should be prepared to control his vehicle if the automation can't handle complex driving situations or if there is a malfunction in the car's automation.

Level 4 means that the car is fully automated. It can handle all driving tasks, even in the most challenging driving conditions. At Level 4, there is no need for human driver intervention. The driver can take control and steer the car by hand in case of a system malfunction or personal desire for self-driving, but in general, the vehicle acts as a self-driving device. Currently, the only Level 4 self-driving car operating is the Waymo fleet. The Waymo case is still a Level 4 in a narrow sense because it is available only in Phoenix, Arizona, and the suburbs. The system is not entirely autonomous, doesn't function in bad weather, and remote human assistance can guide the automated vehicles in certain conditions [12].

Finally, Level 5 represents an autonomous vehicle without human interaction. At this level, cars are autonomous robots that transport passengers and goods independently.

## B. Teleoperation of self-driving cars

The highest level of automation available for public sale is Level 3, and even its availability is minimal [13]. Migration from a Level 3 car to a fully automated Level 4 or a completely autonomous vehicle of Level 5 is challenging. Teleoperation can fill the technological gap between an autonomous vehicle that is not perfect and needs human assistance on rare occasions and the desired fully robotic system. The target ratio between almost fully automated cars on the road and human teleoperators in a single moment is around 1000 [14]. The remote driver controls the computerized vehicle in case the car's automation malfunctions or an unexpected driving situation that the car can't handle safely. The idea is to allow automation to control the vehicle fleet at 99.9% while the remote drivers will intervene in the remaining 0.1% of the driving hours. The car's teleoperation should receive a high-quality video with minimal delay between the camera and the remote driver's monitor. The round-trip signal delay comprises the car's camera-to-remote monitor delay, the teleoperator response time, and the time required to transmit the steering wheel and pedal commands from the distant driver to the autonomous car. This human-machine closed-loop time should be at most 170 ms [15], leaving much less time for the camera to teleoperator's monitor signal travel.

## C. Use of 4G and 5G cellular networks for teleoperation

Experiment with teleoperation over 4G networks [2][16] shows that the upload and download requirements are adequate in many conditions. However, the network needs to be more robust to enable smooth teleoperation. These experiments with the 4G network suggest that the 5G network may be acceptable for reliable teleoperation. A remote driver should have a continuous HD video of high quality and a maximum delay of the image between the camera sensor and the operator monitor of no more than 100 ms. This short interval also contains the video encoding, decoding, and 5G packet transmission time. Therefore, successful remote teleoperation requires high-quality video transmission with a minimum delay over the 5G network. This task is very challenging since network quality is not uniform, and there are many driving conditions where the network bandwidth degrades significantly. A recent preliminary remote driving experiment with the Carla [17] driving simulator shows that even a G2G delay of 70 ms is noticeable but still allows successful and safe remote driving.

### D. The Cognata driving simulator

This research generated the driving scenarios with the Cognata driving simulator. Cognata produces driving movies for autonomous vehicles through a controlled simulation [18]. In addition, the Cognata simulator can create photo-realistic videos that are close to reality. Photo-realistic driving scenarios are essential for video compression research because their compression results reflect the compression of real-life scenarios rich in fine details. On the other hand, non-photorealistic simulators yield simplistic scenes with low spatial complexity, leading to compressed files of much smaller sizes than actual and natural driving scenes. The user can decide where to place cameras, Lidars, radars, GPS & lane detectors on the car's layout. After performing a driving simulation according to a modified driving scenario, the user can download the response of each sensor given by a contiguous video or a collection of Portable Network Graphic (PNG) non-compressed images. Fig. 1 shows a snapshot of a driving scenario taken from the car's front camera. Fig. 2 shows the ground truth semantic segmentation map according to the pixel classes of the Cognata simulator.

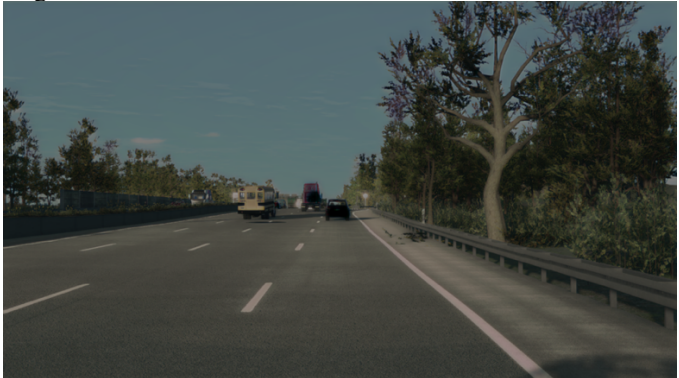

Fig. 1. Snapshot of a driving scenario generated by Cognata in PNG (lossless) format.

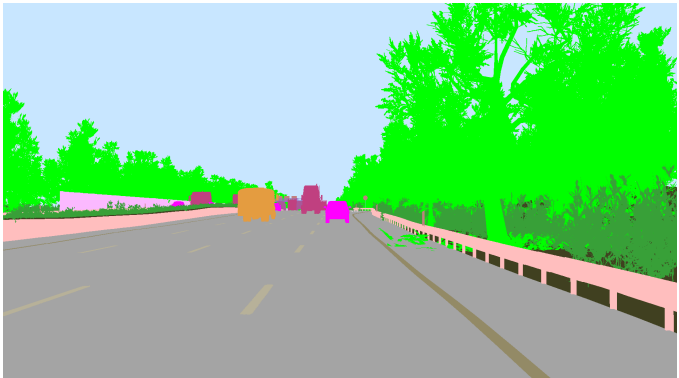

Fig. 2. Semantic segmentation of the original image according to the classes described in Table 1. Every color represents a different class.

### E. Concurrent Video Coder-Encoders (Codecs)

A crucial component of a teleoperation system is the video encoder/decoder. Therefore, selecting an appropriate video compression standard is an important decision. An aging standard may be robust and rich in practical software and hardware implementations but yield poor Rate-distortion (RD) performances. On the other hand, a very recent standard may not be too mature yet and lacks practical and available real-time implementations.

In this research, we chose the HEVC (H.265) [19] over the Advanced Video Coding (AVC, H.264) [20] because of its superior advantage in terms of bit rate saving. The results in [21] show clear benefits of HEVC over AVC and Google's open-source VP9 [22]. The comparative study in [23] compares H.265, H.264, and VP9 for low-delay real-time video conferencing applications. Results show that H.265 is better than H.264 and VP9 by 42% and 32.5%, respectively, in terms of bit rate and for the same PSNR (Peak signal-to-noise-ratio).

Also, contemporary hardware can manage with the HEVC's higher coding complexity. Although there are new compression standards, such as the Versatile Video Coding (VVC)/H.266 [24], AV1 [25], and MPEG5 [26] are recent, and their real-time practical encoding implantations have yet to be made available. Fig. 3, restored from Bitmovin's annual report [27], shows that the AVC is still the most dominant video Codec in current video applications. HEVC is the second. However, if we look at the decoders that developers plan to use in future designs, we can see that the HEVC excels the AVC.

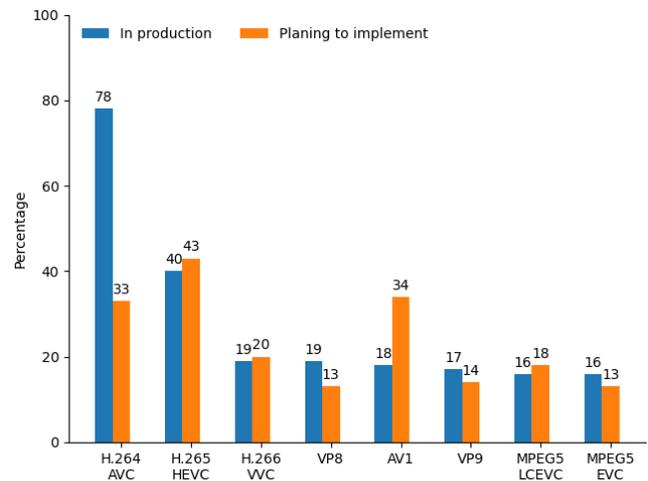

Fig. 3. Developers using a particular video codec – blue, how many plans to use a specific codec in the next 12 months (Total sums to more than 100% because some developers implement more than one codec).

The H.265 encoder partitions the image into CTUs. A CTU is the basic processing block. Its size can be up to 64x64 pixels [19]. The encoder can recursively subdivide each CTU into Coding Units (CU) ranging from 32x32 to 8x8 pixels. Each QG (Quantization Group) has the CU size according to the CTU partitions and can be assigned a separate Differential $QP$ ($\delta QP$) [19] where the $\delta QP$ is the difference from the baseline $QP$. Varying the $QP$ controls the compression quality. The standard defines and allows such fine control that each CU has a separate $\delta QP$. However, for simplicity, in this work, we'll limit the compression quality granularity to the size of a whole CTU.

### F.  Rate Control (RC)

RC is a lossy compression optimization method that fits the compressed video size into a bandwidth constraint. The target is to minimize the video distortion ($D$) at a given encoder rate ($R$) measured in bits/sec. $D$ is the error between the original and reconstructed images from its lossy compressed representation. The RC algorithm fits the compressed video size into the band limitation of the Network. The most common measure for $D$ is the Mean Square Error (MSE). There is always an RD tradeoff. As the Network allows higher $R$, it achieves smaller $D$ values. RC methods have changed significantly during the long journey to attain modern video compression since the '80s. With the development of new video encoding standards, rate control methods were also adjusted for the recent changes and requirements.

In [28], the refresh rate, $QP$, and movement detection threshold are adjusted, regulating the very coarse bitrate of the early H.261 standard. The authors of [29] proposed a quadratic relationship between $QP$, and $R$ for the MPEG2 and MPEG4 compression standards:

$$R = \frac{a}{QP} + \frac{a}{QP^2} \qquad (1)$$

Where a and b are calculated based on the encoding of previous frames.

The authors of [30] showed that the $R$-$\lambda$, RC method proposed for HEVC outperforms previous HEVC RC, which is based on $R$-$Q$ model dependencies and is part of the HM reference software.

The Lagrange multipliers help us select an optimal point along the RD monotonically decreasing convex function [31]. It uses the Lagrange multipliers optimization, where $\lambda$ the Lagrange multiplier depends on frame complexity. The granularity of Li's algorithm can be an entire GOP, a unit of time, at the frame or CTU levels. It can assign a separate $QP$ to every CTU. The Joint Collaborative Team on Video Coding (JCT-VC) included this method in the HEVC Test Model. The $R$-$\lambda$ is implemented in Kvazaar as the default RC method, and we'll review it.

The objective of RC is to minimize the overall $D$ at a target bitrate, $R$:

$$\min_{\{r_i\}_{i=1}^{M}} D = \sum_{i=1}^{M} d_i \quad s.t. \sum_{i=1}^{M} r_i \leq R \qquad (2)$$

Where $d_i$ and $r_i$ are the distortion and target bits for the $i'th$ CTU. $M$ is the total number of CTUs in a frame. The Lagrange method converts the constrained equation in **Error! Reference source not found.**) into an unconstrained optimization equation. Equation (3) can be solved by taking the derivatives with respect to $r_i$. and comparing them to zero. Li et al. showed that the hyperbolic model $\lambda$ fits better than the exponential model and can be written by:

$$\lambda = -\frac{\partial d_i}{\partial r_i} = c_i \, k_i \, r_i^{-k-1} = \alpha_i \, r_i^{\beta_i} \qquad (3)$$

Where $\alpha_i$ and $\beta_i$ are determined by the video content. Since equation (3) has two variables, $\alpha_i$ and $\beta_i$, they can't deduced directly. Li has developed an updating algorithm based on the collocated CTU values of $\alpha$ and $\beta$ and $r_i = bpp$ (bits per pixel) by the following equations:

$$\lambda_{comp} = \alpha_{old} \, bpp_{real}^{\beta_{old}} \qquad (4)$$

Were the actual encoded bits for a whole frame or a CTU is known after the encoding is performed. $\alpha_{old}$ and $\beta_{old}$ are the previous frame or colocated CTU of the same level.

$$lr = ln \frac{\lambda_{real}}{\lambda_{comp}} \qquad (5)$$

Where $lr$ is the log ratio between the estimated $\lambda$ from the allocated $bpp$ to the actual $\lambda$ after the encoding is performed and the actual bpp is known.

Then, from the log ratio, the $\alpha$ and $\beta$ parameters that define the updated $\lambda$ according to the video complexity are updated by:

$$\alpha_{new} = \alpha_{old} + \delta_\alpha \, lr \, \alpha_{old} \qquad (6)$$

$$\beta_{new} = \beta_{old} + \delta_\beta \, lr \, \beta_{old} \, ln \, bpp_{real} \qquad (7)$$

And $\delta_\alpha = 0.1$, $\delta_\beta = 0.05$ are constants that govern the updating rate.

The authors of [32] propose an Optimal Bit Allocation (OBA) RC method. According to the OBA, the method proposes an RD estimation instead of the R–λ estimation for better $D$ minimization. The authors argue that the original R–λ optimization method proposed for HEVC in [33] is not optimal and that the direct calculation of the Lagrange multiplier yields higher distortion. With this RD method, an accurate estimate of $QP$ achieves the RC target.

The method uses the Taylor expansion method to obtain a closed-form solution by iterating over the Taylor expansion. The iterative procedure converges fast, requiring no more than three iterations.

### G.  ROI compression

ROI compression is a lossy video technique that favors certain image areas over others. The research described in [34] implements skin tone filtering to distinguish between humans in a conversational video and background. The video encoder adaptively adjusts the coding parameters such as $QP$, search range, candidates for mode decision, the accuracy of motion vectors, block search at the Macro Block (MB) level, and the relative importance of each MB. Finally, the encoder assigns more resources, bits, and computation power to the MB that resides inside an ROI and fewer resources to the MB that lives inside a Non-ROI.

An RD optimization for conversational video detecting faces' key features allows separation between the human faces and the background [35]. The faces should be inside the ROI. Then, the Largest Coding Units (LCU) that contain parts of faces are searched deeper during the LCU encoding process and assigned lower $QP$. Non-ROI LCUs are searched shallower and assigned higher $QP$ values. The model uses a quadratic equation that finds the relationship between the desired $QP$, the bit budget, the Mean Average Difference (MAD) between the original LCU and the predicted LCU, model parameters, and the weighted ratio between the bits allocated for ROI and the bits assigned for non ROI.

A novel ROI compression for the H.264 SVC (Scalable Video Coding) extension is proposed and studied [36]. ROI is dynamically adjustable by location, size, resolution, and bitrate. The authors of [8] allocate the Lidar bitrate budget according to ROI and non-ROI, significantly improving the quality of self-driving car depth maps. In [37], the encoder allocates the CTU bits according to their visual saliency. Then, the $QP$ for each CTU is calculated by the Lagrange multipliers.

An object-based ROI by a non-supervised saliency learning algorithm for saliency detection is proposed in [38]. This work detects salient areas in two steps: detection and validation.

The authors of [39] propose ROI compression for static camera scenes by considering ROI as blocks significantly different from the previous frame and applying the Discrete Chebyshev Transform (DTT) instead of the conventional Discrete Cosine Transform (DCT).

The authors of [40] propose a DL end-to-end CNN network for rate-distortion optimization (RDO) of a still image JPEG2000 compression. ROI is optimized using a PSNR loss function for the ROI and SSIM (Structural Similarity Index Measure) for the entire image. In [40], ROI rate control for H.265 video conferencing is solved by applying ideas from game theory. Then, the encoder allocates the ROI and non-ROI CTUs using the Newton method.

### H. HEVC Codec applications

The most popular HEVC Codec for research purposes is the HEVC Test Model (HM) reference software developed by the Joint Collaborative Team on Video Coding (JCT-VC). It is a reliable source for ensuring compatibility and adherence to the HEVC standard. Although its time performance improved from the HM-16.6 to the HM-16.12, the encoding must be faster and is still far from real-time requirements [41]. The x265 is among the most popular HEVC Codec. It is continuously developed and improved by MulticoreWare and used or supported by many popular video applications such as FFmpeg, Handbrake, Amazon Prime Video, and many more. It also has fast coding modes suitable for real-time applications. However, providing a $\delta QP$ map for all image Coding Tree Units (CTUs) during encoding for each image is unavailable.

We chose the Kvazaar [42] HEVC Codec for this research. Kvazaar is an open-source academic video codec specializing in high-performance, low-latency encoding using the HEVC (H.265) standard. It's suitable for real-time applications and offers a command-line interface and an API for flexibility. Kvazaar enables the definition of a $\delta QP$ ROI map per image and CTU.

Teleoperation demands minimal G2G delay, necessitating low-delay video encoding. This can be achieved by limiting the encoded video GOP sequence to only Intra (I) and Predictive (P) frames. The highly coding efficient bi-directional (B) frames are not allowed here since their encoding process introduces an additional delay of several frames.

### II. CLOSELY RELATED RESEARCH WORKS

The following thesis [43], explores the possibility of reducing data transmission by compressing a video stream based on the importance of specific features by using static and dynamic feature selection, with more critical components receiving less compression by segmenting the image into ROI and region of non-interest (RONI).

In her Ph.D. dissertation [56] from 2005, Lin proposed ROI rate control encoding for video conferencing with H.263 standard. First, the face in a video conversation is segmented from the image by skin color and the mosaic rule in the $I$ frames, followed by motion vectors only for the $P$ frames. Then, after the segmentation maps are created, ROI quadratic RC methods are applied to the H.263 video compression.

Another work that deals with video bandwidth saving for Autonomous Vehicles (AV) is Bagwe's master thesis [44]. His work reduces the bandwidth required for video transmission between the AV and connected vehicles by skipping redundant frames.

The authors of [45] present two techniques to improve video quality for the remote driver. The first scheme focuses on adaptation of traffic-aware multi-view video stream, beginning with traffic-aware view prioritization. This approach determines the importance of individual camera views based on the current driving situation. The second scheme reduces the bandwidth of a separate camera stream by applying dynamical elliptical masks over the ROI and blurring the area outside the ROI. This causes the video encoder to use fewer bits for the background compressor image blocks.

The authors of [46] address rate assignment in a distributed tile encoding system in multi-resolution tiled streaming services based on the emerging VVC Standard. A model for rate assignment is derived based on random forest regression using spatio-temporal activity and encodings of a learning dataset.

### III. THREE CATEGORIES OF ROI COMPRESSION METHOD

In this work, we categorized the pixel classes into three separate categories. Category 2 contains the classes that are most important for the teleoperation of self-driving cars. These are the traffic signs (standard and large) and traffic lights. During video compression, our goal is to compress CTUs containing pixels of such classes with pixel counts that exceed a certain threshold to the best video quality possible with our constraints of a specific low bandwidth and requirement for ultra-low delay. These classes are considered here as strong ROI. Category 1 contains all the pixel classes that are also important for driving. Among them are the roads, lane marking of various types, cars, pedestrians, and bicycles. In our process of ROI compression, we don't want to degrade the quality of these objects compared to standard compression. The third, Category 0, is the background classes. These include the Sky, Trees, Vegetation, and Buildings. These classes are least important for remote driving, and we can allow us to transmit them with lesser quality.

We use the Cognata [9] proprietary driving simulator to generate driving scenarios. Among the many sensors and maps that this simulator can simulate are the semantic segmentation maps of each simulated video frame. Since the photo-realistic video image sequence and the related semantic segmentation images are generated by the same tool, the semantic segmentation maps are the perfect ground truth of their photo-realistic counterparts.

The Cognata semantic segmentation has around 50-pixel classes, where every class has its unique color. Fig. 1 and Fig.

2 show a photo-realistic image taken from a simulated driving video and its semantic segmentation representation, respectively. For example, the sky pixels are colored blue, and the road pixels are colored gray. The Cognata classes are based on earlier works. Cityscape is notable for scene understanding and semantic segmentation for self-driving cars [47]. It comprises an extensive collection of urban street scenes captured from different cities, with pixel-level annotations that label various objects and areas within the images. The dataset is frequently used to train and evaluate algorithms designed to understand and interpret the visual information present in urban environments. Cityscape has 30-pixel classes.

The division of the classes into three ROI categories that we used in this analysis is described in **Error! Reference source not found.**. It is only a specific proposal tailored to a particular driving environment. It may be varied in a different environment. For example, suppose there is no separation, such as a sidewalk between the building and the road. In that case, it may be more practical to move the building pixel class from the background category to the weak ROI category, or if the acoustic wall is always well separated from the road, moving its class from the background to the weak ROI may be more practical.

Fig. explains the encoding process. (a) describes a non-compressed 'png' image as taken from a photo-realistic simulated driving scenario. The frame rate is 30 fps, and the resolution is HD of 1080x1920 pixels. Image (b) is the semantic segmentation map of (a). Each of the pixel classes has its unique color. (c) describe the conversion of the semantic map from pixel classes into pixel categories according to **Error! Reference source not found.**. All background pixels that belong to the background classes are colored black, all pixels that belong to the weak ROI classes (classes that are important for teleoperation and road understanding) are colored gray, and the pixels that belong to the strong ROI contain the traffic signs and traffic lights which are most important for driving are colored white. According to the HEVC standard [48], the image is divided into CTU rectangles of constant size. In our case, the 1080x1920 HD image has 17x30 CTUs, each of 64x64 pixels. The last row of CTUs is 56x64 pixels in size. Fig. (d) describes the conversion from pixel categories into CTU assignments according to the following rule:

$$M_{CTU}^{i,j} = 0 \qquad (9)$$
$$\text{If } N_{CTU}^{i,j,2} > t_2$$
$$M_{CTU}^{i,j} = 2$$
$$\text{Elseif } N_{CTU}^{i,j,1} > t_1$$
$$M_{CTU}^{i,j} = 1$$

Where $M_{CTU}^{i,j}$ is the CTU category assignment for CTU at row $i$ and column $j$. $N_{CTU}^{i,j,c}$ is the number of pixels of a CTU of category $c$. $t_1$ and $t_2$ are the threshold values for categories 1 and 2, respectively.

In this analysis, $t_1$ and $t_2$ were set to 512, which is $1/8$ of the number of pixels in our CTU. We can observe in Fig. (c) that some of the small traffic signs that are colored white are converted into background CTUs because their pixel count of category 2 doesn't exceed the threshold value. This thresholding saves us bandwidth by not encoding with lower $QP$, CTUs that are far away and contain traffic signs, and traffic lights that are barely visible. Also, CTUs with only a few weak ROI pixels are assigned as background CTUs. Thresholding is also essential in real and not simulated video encoding, where a trained DNN produces the semantic segmentation, has segmentation errors, and can neglect a small number of pixels classified erroneously.

The next step is to take the CTU HD 17x30 category map of every image and convert it into $\delta QP$ values. A category 2 CTU is assigned with $\delta QP = -q$, category 1 with $\delta QP = 0$ (no change from the base $QP$), and category 0 with $\delta QP = q$. $q$ is an integer value between 1 and 10. The video encoder doesn't allow for higher values due to stability issues. The higher the $\delta QP$, the higher the DCT/DST quantization, resulting in a coarser image and reduced bandwidth. The non-compressed movie sequence of 'png' images was converted into a YUV420p raw video by FFmpeg [49], the $\delta QP$ maps of 17x30 values, one map for every image was represented in binary, and both raw video and ROI file were fed into the Kvazaar HEVC video compressor. The bitrate was set to 1 Mbs, and the compressor was set to low delay operation. By using these options, the rate control that Kvazaar uses is a frame-level $\lambda$ domain rate control that adjusts the frame $\lambda$ when advancing along the video frames according to frame bitrate undershoot or overshoot. Then, the base frame $QP_b$ is calculated as a function of $\lambda$. The actual $QP$ of each CTU is then calculated by:

$$QP = QP_b + \delta QP \qquad (10)$$

To avoid out-of-range values of $QP$, clipping should be performed to the $QP$ HEVC range:

$$QP = CLIP(0, 51, QP) \qquad (11)$$

The video sequence was also compressed by Kvazaar without an ROI map. In the former case, the rate control that is default applied is an implementation of Li's [30] $\lambda$ domain rate control where the rate control is also optimized over the individual CTUs.

The metrics we use to evaluate the performance of the ROI compression over the three ROI categories are based on the PSNR and SSIM. Since we are interested in the MPSNR (mean PSNR) concerning the three different ROI categories, we'll define the MPSNR by the following:

$$MPSNR^i = \frac{1}{N^i} \sum_{\substack{Over\ all\ CTUs \\ PSNR\ of \\ category\ i}} PSNR_{CTU}^i \qquad (12)$$

Were $N^i$ is the number of CTUs of category $i$ and $PSNR_{CTU}^i$ is the PSNR of a CTU of $i$ category. Similarly, we'll define the categorical MSSIM (Mean Structural Similarity Index). by the following:

$$MSSIM^i = \frac{1}{N^i} \sum_{\substack{Over\ all\ CTUs \\ MSSIM\ of \\ category\ i}} MSSIM_{CTU}^i \qquad (13)$$

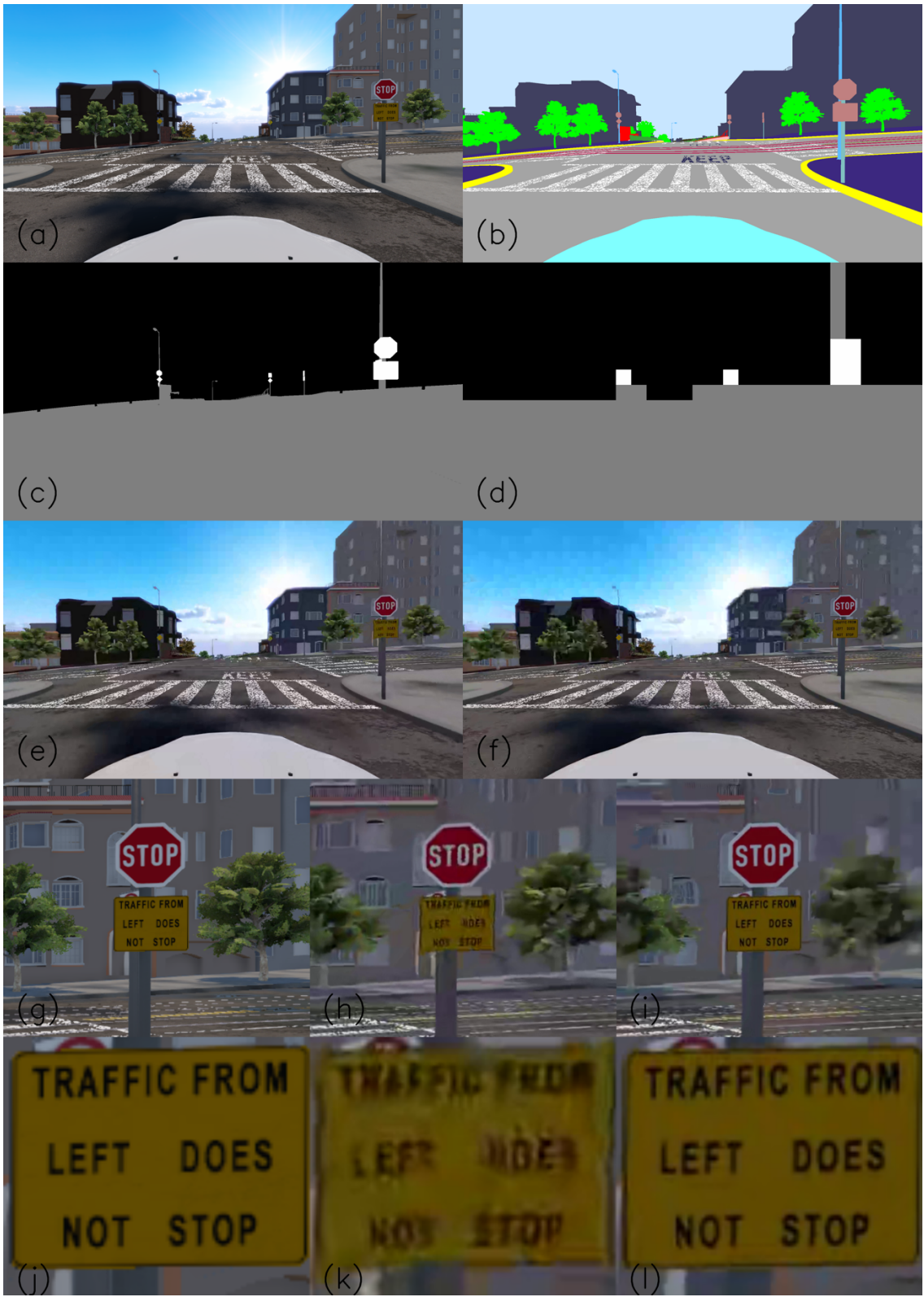

Fig. 4. Description of the three categories ROI video compression method. Video bandwidths are 1 Mbps. (a) is the original non-compressed image. (b) is the semantic segmentation of the original image. (c) is the categorization of the semantic classes. (d) is the CTUs assignments. (e) is the standard compressed image. (f) is the ROI compressed. (g) – (i) are the enlarged traffic signs area as appeared in the original, standard, and ROI compressed images. (j) – (l) are enlargements of the textual traffic sign.

Table 1 Categorization of the image segmentation classes into 0 (background), 1 (weak ROI), and 2 (strong ROI) pixel categories.

| Categorization of pixel segmentation classes according to their importance for remote driving | |
|---|---|
| Category 2 (most important, Strong ROI) | Traffic sign, Large sign, Traffic light |
| Category 1 (important, Weak ROI) | Road, Parking, Railway, Lane marking, Road marking shape, Sidewalk, Biking lane, Car, Truck, Bus, Motorcycle, Drone, Rider, Pedestrians, Bird, Curb, Tunnel, Bridge, Fence, Guardrail, Acoustic wall, Props, Electricity cable, Pole, Electric light pole, Ego Car, Bicycle, Van, Lane Line Types: Dashed, Solid, Double Solid, Double Dashed, Double Solid and Dashed, None Animal, Gantry, Trailer, PersonalMobility, Construction Vehicle, Rock |
| Category 0 (Background Classes) | Unlabeled, Tree, Low vegetation, Sky, Building, Building Far |

## IV. SIMULATION OF DRIVING CASE STUDIES

### A. Analysis of a simulated driving movie clip containing traffic signs as well as textual signs

The following comparison between the standard Kvazaar and ROI compression is for a Cognata-simulated driving movie clip from Lombard Street in San Francisco. Fig. 4(a) is an image taken from the simulated movie clip. (b) describe its semantic segmentation result were each class is uniquely colored. (c) describe the categorization of the pixel classes into three categories as described in Table 1. The white is for Category 2, the gray is for Category 1, and the black is for Category 0. (d) describe CTU thresholding according to equation (9). Both compression results are shown in Fig. (e) and (f). (e) is an image taken from a compressed video sequence by standard compression, and (f) is from the ROI compression. Our observations reveal a notable enhancement in the visual quality of the textual sign displaying the message "TRAFFIC FROM LEFT DOES NOT STOP" with ROI compression. Additionally, there is a discernible improvement in the visual clarity of the Stop sign. It is worth noting that common traffic signs are engineered to maintain visibility even under adverse conditions such as inclement weather, wear and tear, and deliberate damage, characteristics that extend to poor video quality as well. We can observe that the quality of the weak ROI objects, such as roads, road marks, and sidewalks, stays roughly the same and even improves slightly. At the same time, there is a significant degradation in the quality of the background objects. The perceived sky quality in the standard compressed video is much better than the ROI compression. Meanwhile, the sky, buildings, and trees appear more blocky than in the standard video. (g), (h), and (i) show the textual and stop signs for the original non-compressed, standard, and ROI compressed. Images (j), (k), and (l) are the enlargements of the textual sign. A remote driver that observes this textual sign by standard video can hardly read the sign, while by applying ROI compression, the sign is well readable. Fig. describes traffic signs taken from different frames from the same driving video simulation. Column (a) shows the non-compressed traffic signs taken from the PNG images. Column (b) describes the compression results by the standard method, while column (c) describes the traffic signs that appear after the ROI compression. The traffic signs in column (c) seem much closer to their original images than their standard compression images.

Fig. and Fig. compare image quality metrics between the three ROI categories of the standard and ROI compression. Fig. compares the ROI categorical PSNR of both video compression methods. PSNR of every category is calculated by averaging over all of its individual CTU PSNR of the same ROI category. The length of the movie sequence was 600 images, but for clarity, only part of the sequence is shown here. The means of the metric are calculated over the whole image sequence. The dashed lines describe the PSNRs for the three categories of background, weak ROI, and strong ROI, and the contiguous lines represent the three categories of PSNRs for the ROI compression. We can observe that the blue lines that describe the background PSNR move on average 3 dB downward when switching from the standard to ROI compression, leaving more bandwidth in the strong ROI category. This change can also be reflected when comparing the sky, trees, and building background categories, as shown in **Error! Reference source not found.**(e) and (f), where there is a significant degradation in background quality. Comparing the yellow lines shows a slight improvement in the weak ROI by applying ROI compression. In this case, it is 0.7 dB, but in general, we don't want to degrade the quality of the weak ROI that contains classes that are important for remote driving, but any figure that is not negative can be acceptable. The exact figures above show that the road quality, road marks, cars, pedestrians, and the classes important for safe driving remain about the same or slightly improved. Comparing the two red lines shows the improvement in the strong ROI CTUs. In this example, the gain is more than 5 dB and is very noticeable when looking again at Fig. (e) and (f).

Fig. compares both compression methods' categorical Mean Structural Similarity Index (MSSIM). The MSSIM is a variation of SSIM that calculates the mean value of SSIM across local image regions, providing a more global measure of image similarity [50]. Here, we average the SSIM of all CTUs of a specific category. Similar to the PSNR explanation in Fig. . The MSSIM metric for the background (blue dashed and contiguous lines) degrades from the mean over all images of 0.85 to 0.79, the weak ROI improves slightly from 0.75 to 0.77, and the strong ROI that contains the traffic signs improves significantly from 0.75 to 0.9 enabling easy reading of the textual traffic sign during driving.

As expected, the overall quality metrics for the video sequence that is ROI compressed should be lower than that of standard compression, which doesn't prioritize any region in the image. Fig. 8 shows the movie clip images PSNR in blue for standard compression and yellow for ROI compression. The standard compression PSNR is almost always higher than the ROI compression since the optimization of the overall PSNR can be achieved by minimizing the error variance among the CTUs. When comparing the average PSNR of the standard compression to that of the ROI compression, we lose 1.1 dB. Fig. shows the video SSIM for both standard and ROI compression. Again, the image metric is degraded. SSIM is reduced from 0.8 for the standard compression to 0.78 for the ROI compression. This comparison is essential as a preliminary logical check that the ROI compression method should degrade the overall metrics of the video motion sequence. However, there is a considerable benefit in increasing the quality of the ROI, predominantly the strong ROI.

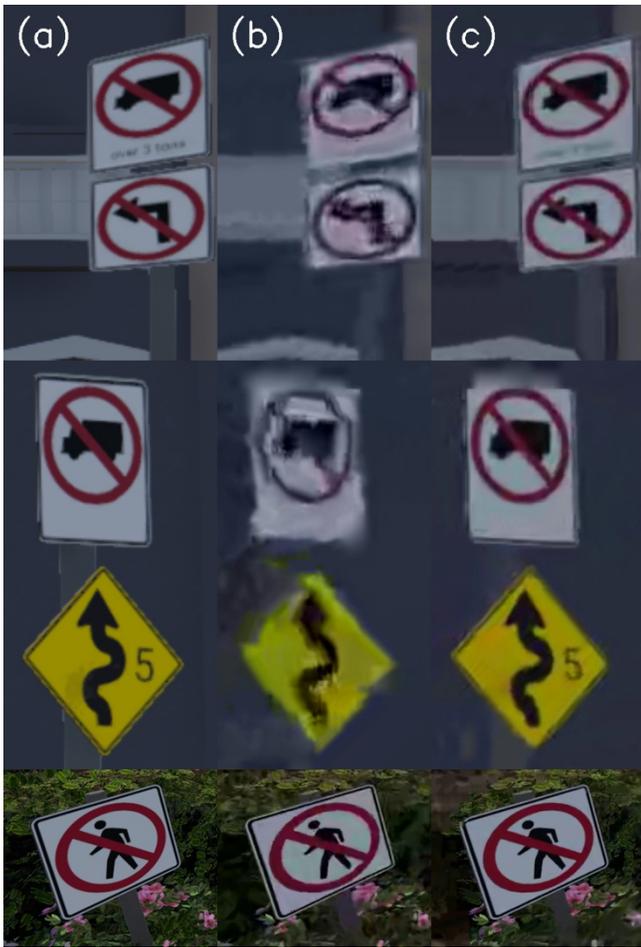

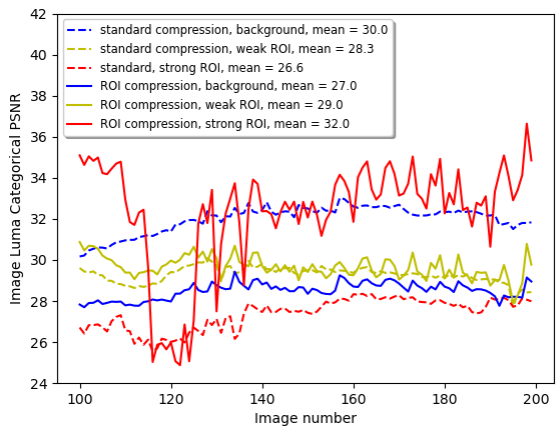

Fig. 5. Traffic signs were taken from the same simulated driving video but from other frames. (a) describes traffic signs taken from the non-compressed images. (b) describes traffic signs as compressed by standard method. (c) describes the traffic signs that are compressed by the ROI method.

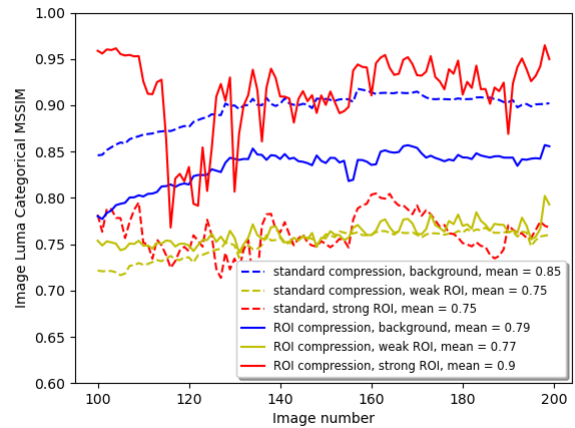

Fig. 7. The Categorical image sequence luminance MSSIM. The dashed lines describe the standard compression background, weak ROI, and strong ROI, while the contiguous lines describe the ROI compression MSSIMs.

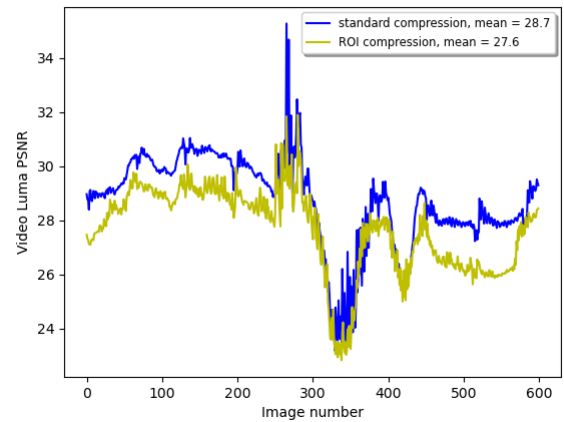

Fig. 8. The overall video luminance PSNR for standard and ROI compression.

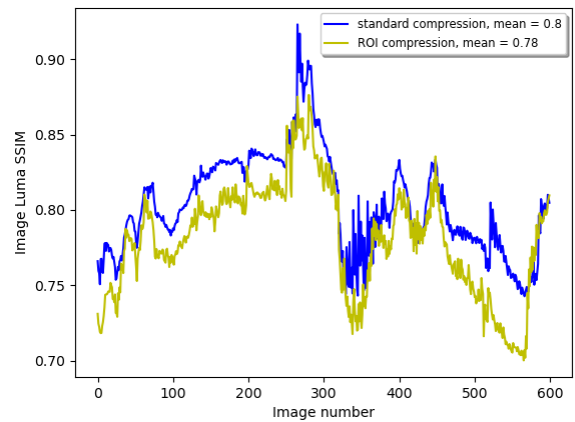

Fig. 9. The overall video luminance SSIM for standard and ROI compression.

Fig. 6. The Categorical image sequence luminance PSNR. The dashed lines describe the standard compression background, weak ROI, and strong ROI, while the contiguous lines describe the ROI compression PSNRs.

*B.     Analysis of a simulated driving movie clip containing traffic lights*

The second example is taken from a simulated driving clip in Munich, Germany, and contains also traffic lights. It is described in Fig. 5. Image (a) illustrates the non-compressed image with traffic lights. Image (b) is the semantic segmentation of the original image. The traffic signs are pinkish-orange, and the lights are pale purple or Lavender. (c) Describe the categorization of every pixel to the background (black), weak ROI (gray), and strong ROI (black) according to Table 1. (d) shows the result after applying the thresholding as described in eq. (). (e) and (f) are the standard and ROI compression results. Again, the sky, buildings, and trees degrade in quality, making their images blurry and bulky. The weak ROI of roads, cars, and sidewalks remains at least with their standard compression quality. The traffic lights, as well as traffic signs, significantly improve their visible quality. (g) to (i) describe the enlargement of the traffic light with a 'Front and right' traffic sign above and 'Main Road' traffic sign below. Again, the traffic lights and signs are transmitted with much better quality on account of the background's bandwidth, which contains the sky, buildings, and low vegetation that are less critical for the remote driver.

Fig. 4 describes the categorical PSNR comparison between the standard and ROI compression. The red lines represent an improvement of 6.5 dB on average in the CTUs containing traffic lights or traffic signs. There are missing values in the red curves because not all images have traffic signs or traffic lights. Also, CTUs containing strong ROI class pixels should pass the threshold pixel count as defined in Eq. (9). The yellow lines, describing the weak ROI, show almost no change, and the blue lines indicate a 3 dB reduction.

Fig. 6 shows a similar comparison but for the categorical MSSIM metric. The strong ROI improves on average from 0.75 to 0.9, the weak ROI improves slightly from 0.75 to 0.77, and the background reduces from 0.85 to 0.79.

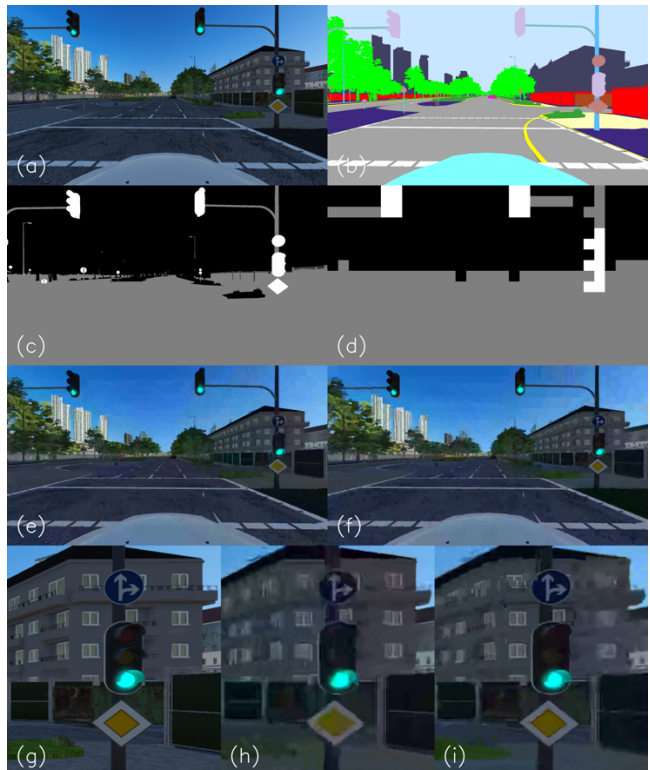

Fig. 5. Description of the three categories of ROI video compression methods at 1 Mbps (a) is the original non-compressed image. (b) is the semantic segmentation of the original image. (c) is the categorization of the semantic classes. (d) is the CTUs assignment. (e) is the standard compressed image. (f) is the ROI compressed. (g) – (i) are the enlarged traffic lights and traffic signs surroundings as appeared in the original, standard, and ROI compressed images.

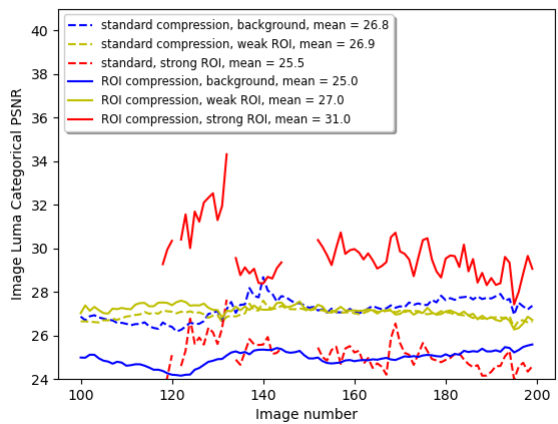

Fig. 4. The Categorical image sequence luminance PSNR for the Munich driving video clip. The dashed lines describe the standard compression background, weak ROI, and strong ROI, while the contiguous lines describe the ROI compression PSNRs. There are missing values in the red curves because not all images have traffic lights or signs of significant size.

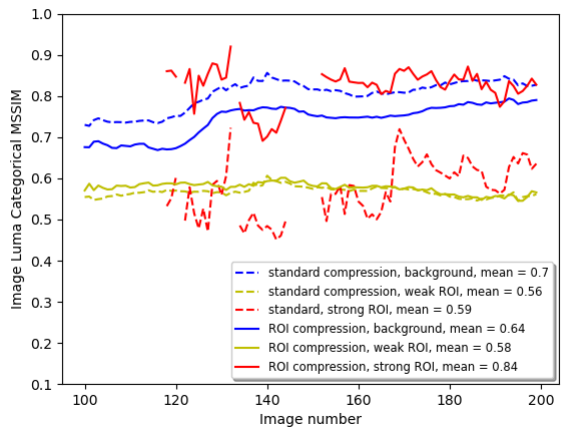

Fig. 6 The Categorical image sequence luminance MSSIM. The dashed lines describe the standard compression background, weak ROI, and strong ROI, while the contiguous lines describe the ROI compression MSSIMs.

## VI. Compare the three ROI categories with the two categories method

The clear advantage of a three-category ROI compression over only ROI and background for the visibility of traffic signs is shown in this section. The visibility of a textual traffic sign is compared among the compression methods described in this work. The movie clip we analyze is the same as defined in section IV. All compressed movie clips from the images are compressed to 1 Mbps bandwidth. Fig. 7(a) and (b) show the same image from two and three ROI compression methods. The general quality of both images is similar except for the quality of the traffic sign area, which significantly improves. Rows (c), (d), and (e) describe the traffic sign along the driving path at three different locations.

The four columns of (c) to (e) rows describe (from left to right) the original non-compressed 'PNG' images, standard compression, two ROI, and three ROI compression. Results show that categorizing the image classes into only background and ROI is far from enough to make the textual traffic sign readable. Separating the original ROI into at least weak and strong ROI is necessary.

Table 2 shows the PSNR and MSSIM metrics for comparing the three compression methods. Metrics are calculated by averaging all images in the movie clips. As expected, the overall metrics degrade by more than 1 dB for the PSNR and 0.02-0.03 for the MSSIM, the background quality degrades significantly, and the weak ROI improves slightly for both methods. In contrast, in the strong ROI region, there is a slight improvement for the two categories of ROI compression and a vast improvement for the three categories method.

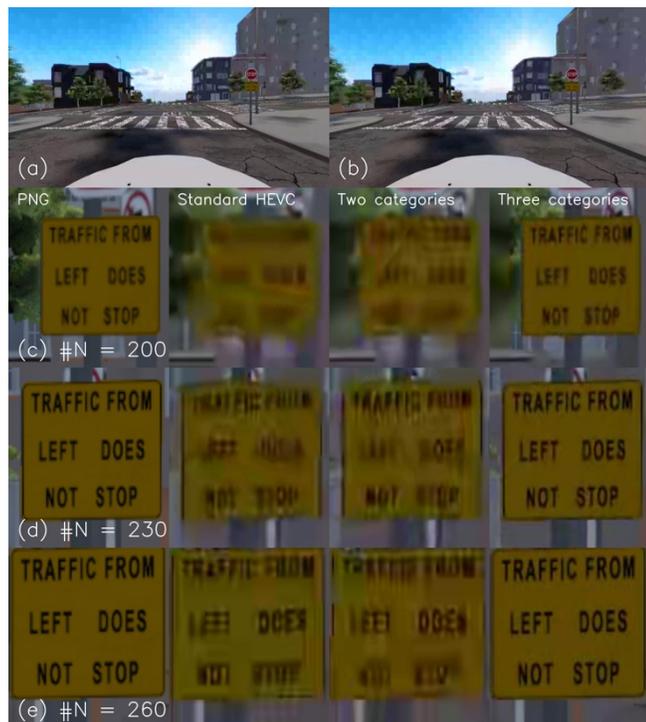

Fig. 7 Comparison of the video encoding methods. (a) and (b) describe the tree and two categories of ROI compression results. The four columns from left to right are the original PNG non-compressed images, the standard HEVC compression results, the two categories of ROI, and the three categories of ROI.

Table 2 PSNR comparison between standard, three, and two ROI video encoding.

| | Overall PSNR | PSNR per category | | |
|---|---|---|---|---|
| | | Background | Weak ROI | Strong ROI |
| Standard | 28.7 | 30.0 | 28.3 | 26.6 |
| Two categories | 27.4 | 26.6 | 29.3 | 27.3 |
| Tree categories | 27.6 | 27.3 | 28.9 | 33.1 |

Table 3 MSSIM compares standard, three, and two ROI categories of video encoding.

| | Overall MSSIM | MSSIM per category | | |
|---|---|---|---|---|
| | | Background | Weak ROI | Strong ROI |
| Standard | 0.8 | 0.85 | 0.75 | 0.75 |
| Two categories | 0.77 | 0.77 | 0.78 | 0.77 |
| Tree categories | 0.78 | 0.79 | 0.77 | 0.9 |

## VII. ANALYSIS ACROSS VARIOUS BANDWIDTHS AND Q VALUES

We analyzed three bandwidths (1, 2, and 4 Mbps) and three q values (3, 5, and 10) to assess the visibility enhancement for traffic signs and lights. Actual bandwidths are also provided alongside the target values to ensure the bandwidth is consistently maintained, even for ROI compression. The results of the PSNR/MPSNR and SSIM/MSSIM are described in Table 4 and Table 5, respectively. The rows representing the visibility of traffic signs and lights designated with $q$ as not applied (NA) describe the default Kvazaar rate control results without using ROI techniques. Comparing the MPSNR for the strong ROI shows significant improvement for all bandwidths. Up to 6.5 dB for the 1 Mbps, 7.6 dB for the 2 Mbps, and 8.4 dB for the 4 Mbps. The quality of the weak ROI never degrades and even improves up to 1 dB depending on bandwidth and $q$. Improvement in ROI quality is due to the migration of bandwidth budget from the background, which loses up to 1.3 dB depending on bandwidth and $q$.

Table 4. PSNR and MPSNR analysis for the Lombard driving clip at various bandwidth and q values. NA is for not applied.

| Band Width (Mbps) | | q | Overall PSNR [db] | MPSNR per category [db] | | |
|---|---|---|---|---|---|---|
| Target | Actual | | | Background | Weak ROI | Strong ROI |
| 1 | 1.008 | NA | 28.7 | 30 | 28.3 | 26.6 |
| 1 | 1.016 | 3 | 28.7 | 29 | 29 | 28 |
| 1 | 1.01 | 5 | 28.5 | 28.9 | 28.9 | 29.8 |
| 1 | 1.006 | 10 | 27.6 | 27.3 | 28.9 | 33.1 |
| 2 | 2.008 | NA | 30.3 | 31.5 | 29..8 | 28.4 |
| 2 | 2.012 | 3 | 30.3 | 30.9 | 30.4 | 30.7 |
| 2 | 2.012 | 5 | 30.1 | 30.2 | 30.6 | 32.3 |
| 2 | 2.006 | 10 | 29.1 | 28.5 | 30.7 | 36 |
| 4 | 4.003 | NA | 32.1 | 33.1 | 31.5 | 30.3 |
| 4 | 4.008 | 3 | 32 | 32.4 | 32.2 | 33.1 |
| 4 | 4.004 | 5 | 31.8 | 31.8 | 32.4 | 34.8 |
| 4 | 3.995 | 10 | 30.8 | 29.9 | 32.6 | 38.7 |

Table 5. SSIM and MSSIM analysis for the Lombard driving clip at various bandwidth and q values.

| Band Width (Mbps) | q | Overall SSIM | MSSIM per category | | |
|---|---|---|---|---|---|
| | | | MSSIM Background | MSSIM Weak ROI | MSSIM Strong ROI |
| 1 | NA | 0.8 | 0.85 | 0.75 | 0.75 |
| 1 | 3 | 0.81 | 0.84 | 0.77 | 0.82 |
| 1 | 5 | 0.8 | 0.82 | 0.77 | 0.86 |
| 1 | 10 | 0.78 | 0.79 | 0.77 | 0.93 |
| 2 | NA | 0.84 | 0.85 | 0.75 | 0.75 |
| 2 | 3 | 0.84 | 0.83 | 0.77 | 0.87 |
| 2 | 5 | 0.84 | 0.86 | 0.82 | 0.92 |
| 2 | 10 | 0.82 | 0.82 | 0.82 | 0.96 |
| 4 | NA | 0.88 | 0.92 | 0.84 | 0.88 |
| 4 | 3 | 0.88 | 0.91 | 0.85 | 0.93 |
| 4 | 5 | 0.88 | 0.89 | 0.86 | 0.95 |
| 4 | 10 | 0.86 | 0.85 | 0.86 | 0.98 |

## VIII. LIMITATIONS

ROI compression is effective only if there is enough bandwidth to be transferred from the background area to the ROI. Therefore, the background should be a substantial portion of the image and considerable complexity. This is not always true, but in many cases, the vehicle's camera view contains significant background areas, such as the sky and vegetation in rural areas or buildings in urban areas.

Separating the ROI into multi-categories of importance or at least into weak and strong ROI relies on the condition that the strong ROI area is small. This may not be true in crowded urban areas with many traffic lights and signs. This method is effective when the bandwidth allocated for the CTUs that contain the traffic signs and traffic lights can be increased by an order of magnitude, and this is possible in many cases where the image includes a few close traffic signs and lights.

This research relied on artificial movie clips from a photorealistic driving scenarios simulator. The semantic segmentation ground truth was also available. We must use non-compressed video sources and real-time DL semantic segmentation methods in a realistic environment, not a simulated one. Such systems exist today. For example, the SOTA real-time image segmentation technique described in [51]. Its small PIDNetS model can reach 90 FPS with the Nvidia 3090 GPU at an mIOU (mean intersection over union) accuracy of around 79%. The deficiencies of the practical semantic segmentation methods from the ground truth can be mitigated by applying our CTUs ROI assignment thresholding method.

## IX. CONCLUSIONS

In summary, this research has focused on refining and evaluating a region of interest (ROI) compression method within the context of autonomous driving scenarios. We utilized the Cognata proprietary driving simulator to generate realistic driving scenarios, with access to ground truth semantic segmentation maps. The segmentation maps contained approximately 50-pixel classes, each uniquely colored, representing various elements in the simulated environment. We categorized these pixel classes into three distinct ROI categories: strong ROI, weak ROI, and background. Strong ROI included crucial elements like traffic signs and lights, while weak ROI encompassed important driving features such as roads, lane markings, vehicles, and pedestrians. Background classes represented less critical elements like the sky, trees, vegetation, and terrain. Our approach involved applying ROI-based compression, with different quantization levels for each category, to prioritize the quality of critical driving elements while conserving bandwidth. The process involved converting pixel categories into Coding Tree Unit (CTU) assignments based on pixel count thresholds, ensuring that strong ROI elements received the highest compression quality. We conducted an extensive analysis using simulated driving scenarios. We evaluated the image quality metrics of PSNR and MSSIM for the different ROI categories. The results revealed significant improvements in the quality of the strong ROI elements, particularly traffic signs and traffic lights, while at least maintaining the quality of the weak ROI elements. The new multi-category ROI compression method shows a clear advantage over an ROI compression that partitions the video into only ROI and background.

## ACKNOWLEDGMENT

We want to express our appreciation to Ofer Lapid, Giora Golan, Eitan Efron, and Dani Rosenthal from the Andromeda Consortium to Eli Shapira from DriveU LTD, Omri Reftov from Cognata LTD, and our students Shir Milstein and Daniel Medvedev.